\documentclass[a4paper]{statsoc}

\usepackage[T1]{fontenc}
\usepackage[utf8]{inputenc}

\usepackage{fullpage}
\usepackage{natbib}
\usepackage{algorithm}
\usepackage{graphicx}
\usepackage{algorithmic}

\newcommand{\bx}{\mathbf{x}}
 
\newcommand{\bu}{\mathbf{u}}

\newcommand{\by}{\mathbf{y}}
\newcommand{\bz}{\mathbf{z}}

\usepackage{amscd,amsmath,latexsym,amssymb,amsfonts}
\usepackage{mathrsfs} 
\usepackage{bm}

\usepackage[a4paper,left=2cm,top=2cm,right=2cm,bottom=2cm,ignoreheadfoot]{geometry}

\title{Three discussions of the paper 
`sequential quasi-Monte Carlo sampling', by M. Gerber and N. Chopin, Read before The Royal Statistical 
Society at a meeting organised by the Research Section on Wednesday, December 10th, 2014}
\author{Julyan Arbel\footnote{julyan.arbel@carloalberto.org}, Collegio Carlo Alberto and University of Torino}

\author{Igor Pr\"unster\footnote{igor.pruenster@unito.it}, Collegio Carlo Alberto and University of Torino}

\author{Christian P. Robert\footnote{xian@ceremade.dauphine.fr}, Universit\'e Paris-Dauphine and University of Warwick}

\author{Robin J. Ryder\footnote{ryder@ceremade.dauphine.fr}, Universit\'e Paris-Dauphine}
\begin{document}
\maketitle 

\begin{abstract}
This is a collection of three written discussions of the paper 
``sequential quasi-Monte Carlo sampling" by M. Gerber and N. Chopin, following the
presentation given before the Royal Statistical Society in London on December 10th, 2014.
\end{abstract}

\section{Consequences on Bayes nonparametrics (J. Arbel and I. Pr\"unster)}
We congratulate the authors for a stimulating paper, accessible to laypersons, and thank them for sharing their code.
We would like to comment on the potential implication of SQMC in Bayesian nonparametrics.

Nonparametric mixtures are commonly used for density estimation and can be thought of as an extension of finite mixture
models when the number of clusters is unknown. The setting is as follows: observations $\by_{1:t}$ are spread out into
$k_t$ clusters; the cluster labels, or allocation variables, are denoted by $\bx_{1:t}$, and are interpreted as the
\textit{states} of $\by_{1:t}$ in the context of SMC. Note that the states are discrete and elements of
$\{1,\ldots,k_t\}$, where $k_t$ denotes the number of clusters.
The $n_{t,j}$ observations allocated to the same cluster $j$ are iid from a given parametric kernel $K(\cdot;
\theta_j)$. This model is static in the sense that data are usually not recorded sequentially. However, P\'olya urn type
schemes \citep{blackwell1973ferguson}
essentially formulate this model sequentially by artificially thinking of observation $\by_t$ to occur at time $t$. So
the nonparametric mixture model can be cast as an SMC sampler as follows
\citep[see][]{liu1996nonparametric,fearnhead2004particle}
\begin{align*}
\by_t \vert \bx_t, \theta &\sim K(\by_t; \theta_{\bx_t}),\\
\bx_t \vert \bx_{1:t-1} &\sim f^X(\bx_t|\bx_{1:t-1}).
\end{align*}
The predictive distribution $f^X$ is  particularly simple for the Dirichlet process
\citep{ferguson1973bayesian,blackwell1973ferguson}. Denote by $\alpha$ and $G_0$ the precision parameter and the base
measure of the Dirichlet process. Then the kernel for the states is given by the following probability mass function on
$\{1,\ldots,k_{t-1}+1\}$
\begin{equation}\label{eq:mt}
m_t(\bx_{1:t-1},\bx_t = j) = p_{t,j}\propto
\left\{
\begin{array}{ll}
n_{t-1,j} K_j(\by_t\vert\bx_{1:t-1}) \quad &\text{ if } j\leq k_{t-1}\\
\alpha K_0(\by_t) \quad &\text{ if } j = k_{t-1}+1
\end{array}
\right.
\end{equation}
where $K_j$ and $K_0$ involve integrals of $K$ and $G_0$ which can be calculated in conjugate cases and approximated
otherwise. Note that $m_t(\bx_{1:t-1},\bx_t)$ is also available in closed-form expression for broader classes of random
probability measures including the two-parameter Poisson--Dirichlet process \citep{pitman1997two} and normalised random
measures with independent increments \citep{regazzini2003distributional,james2009posterior}.

As stressed by Gerber and Chopin, the key ingredient of SQMC is the deterministic transform $\Gamma_t$. A possible
choice for \eqref{eq:mt} is
\begin{equation}\label{eq:QMC}
\Gamma_t(\bx_{1:t-1},\bu_t) = \min\Big\{ j\in\{1,\ldots,k_{t-1}+1\}: \sum_{i=1}^{j}p_{t,i} > \bu_t \Big\}
\end{equation}
where $\bu_t\sim\mathcal{U}([0,1))$. The case of discrete proposal $m_t(\bx_{1:t-1},\bx_t)$ does not seem to be
discussed in the paper. Would the authors expect a gain in efficiency by considering a quasi-Monte Carlo alternative
as~\eqref{eq:QMC} instead of standard SMC?

\section{Spectrum of the sqMC method (C. Robert)}
Quasi-Monte Carlo methods are ignored by the statistical community, despite regular
attempts by well-respected researchers to induce a broader use of those methods. The authors 
are to be congratulated and will hopefully contribute to a wider diffusion of qMC methods.

At a purely mathematical level, that randomised low discrepancy sequences produce both unbiased estimators and error
rates of order $\text{O}(N^{-1}\log(N)^{d-1})$ at cost $\text{O}(N\log(N))$ implies that sqMC
methods should {\em always} be replace regular Monte Carlo methods. Since this is not the case, we may
wonder at the reasons. The major difficulty with qMC
stands in its requirement to expressing Monte Carlo estimators as deterministic transforms of a fixed number of uniforms.
This is a strong deterrent to the dissemination of qMC
alternatives, as moving to a random number of baseline uniforms does not seem achievable in full generality.
It is thus no surprise that the proposal appears in connection with particle filters. Indeed, this field centres
on dynamic importance sampling and hence enjoys a local iid-ness that relates better to
qMC integrators than single-chain MCMC algorithms. For instance, each
particular resampling step consists in multinomial simulation, hence could be turned into
qMC. Actually, lower variance solutions that amount to qMC, like systematic and residual sampling
\citep{gundersenjensen87,fearnhead:1998,chopin:2003}, have been proposed in the particle literature.

Here, the authors apply qMC to the particles themselves, using a {\em global} low discrepancy sequence for the
resampling and the move steps: as the cost of using the Hilbert curve grows quickly with dimension, two sequences in
lieu of one would bring a significant cost reduction. Moreover, they still assume the deterministic transform
representation
$$
\mathbf{x}_t^n = \Gamma_t(\mathbf{x}_{t-1}^n,\mathbf{u}_{t}^n)\,,
$$
which is a stumbling block for those contemplating switching to qMC, as illustrated in the paper using only
Normal baseline distributions. In a sequential setting with unknown parameters $\theta$, the transform $\Gamma$ changes
each time $\theta$ is modified, which impacts computing cost when the inverse cdf is not
available. Since simulating $\theta$ is unlikely to benefit from qMC improvements, one may
question the relevance of the overall scheme: the most variable item in the Gibbs sampling steps expands
its inefficiency to the joint kernel.

In connection with Lemma 1 and the sqMC approximation of the evidence, 
Rao-Blackwellisation using all {\em proposed} moves could be considered:
is this easier and significant within qMC since the gain may be of a lower order?
Similarly, the trick at the end of \S 4.2.1, using sqMC on a single instead of $(t+1)$ vectors, is
unclear, but I wonder at a connection with the particle learning literature
\citep{lopes:carvalho:johannes:polson:2010,carvalho:johannes:lopes:polson:2010}.

In conclusion, I am looking forward the aftermath of this Read Paper that will expose whether qMC is
bound to become the reference in simulation computational statistics or to remain a niche activity away
from mainstream simulation.

\section{ABC with sqMC (R. Ryder)}
\newcommand{\btheta}{\boldsymbol{\theta}}

One of the reasons why the authors are so successful in their new methodology is the ability to know in advance the
number of uniform variables which are needed for one step of the algorithm. This allows greater efficiency and, to use
the terminology at the end of section 1.2 of the paper, the use of QMC point sets (fixed length) instead of QMC
sequences (unbounded length). Nonetheless, as noted by the authors, the use of QMC sequences is at least theoretically
possible.

When reading the paper, I was immediately curious about the possibility of adapting this methodology in an ABC
(Approximate Bayesian Computation) setting. In an attempt to develop a naive ABC population Quasi Monte Carlo algorithm,
it is natural to use QMC sequences instead of QMC point sets, as shown in Algorithm \ref{algo:ABCSQMC}, which is adapted
from \cite{MPRR12}.

\begin{algorithm}[t]
\caption{Likelihood-free population Quasi Monte Carlo sampler}\label{algo:ABCSQMC}
\begin{algorithmic}
\STATE Input: observations $\by$, decreasing sequence $(\epsilon_1,\ldots,\epsilon_N)$, prior $\pi$ with cdf $F_\pi$.
\STATE At iteration $t=1$,
\STATE Set $j=1$
  \FOR {$i=1$ to $N$}
    \REPEAT
       \STATE Generate $u_j$ as the $j$th element in an 1 dimensional RQMC sequence.
	\STATE Set  $\btheta_i^{(1)}= F_\pi^{-1}(u_j)$ and $\bz\sim f(\bz\mid \btheta_i^{(1)})$.
	\STATE Increment $j$.
     \UNTIL $\rho(\eta(\bz),\eta(\by))\leq\epsilon_1$
    \STATE Set $\omega^{(1)}_i=1/N$
    \ENDFOR
\STATE Take $\Sigma_1$ as twice the empirical variance of the $\btheta_i^{(1)}$'s
\FOR {$t=2$ to $T$}
  \STATE Set $j=1$
  \FOR {$i=1$ to $N$}
    \REPEAT
      \STATE Generate $(u_j, v_j)$ as the $j$th element in a 2 dimensional RQMC sequence.
	\STATE Select an ancestor $\btheta_i^\star$ from the $\btheta_\cdot^{t-1}$ using $v_j$ as for SQMC
      \STATE Generate $\btheta_i^{(t)}$ using $\btheta_i^\star$ and $u_j$ as for SQMC
	\STATE Increment j.
    \UNTIL $\rho(\eta(\bz),\eta(\by))\leq\epsilon_t$\\
    \STATE Set $\omega^{(t)}_i\propto \pi(\btheta^{(t)}_i)/\sum_{j=1}^N \omega^{(t-1)}_j
	\varphi\left\{ \Sigma_{t-1}^{-1/2}\left(\btheta^{(t)}_i-\btheta^{(t-1)}_j\right)\right\}$
     \ENDFOR
\STATE Take $\Sigma_t$ as twice the weighted variance of the $\btheta_i^{(t)}$'s
\ENDFOR
\end{algorithmic}
\end{algorithm}

I tried this algorithm on a toy example: $y_k\sim N(x_k, \sigma^2)$ where $(x_k)$ is a Markov chains taking values in $\{-2,2\}$, with probability of switching at any iteration equal to $\theta$, and attempted to estimate $\theta$. For the generation of QMC sequences, I used the \texttt{randtoolbox} R package \cite{randtoolbox}.

\begin{figure}[hbt]
\centerline{\includegraphics[width=.4\textwidth]{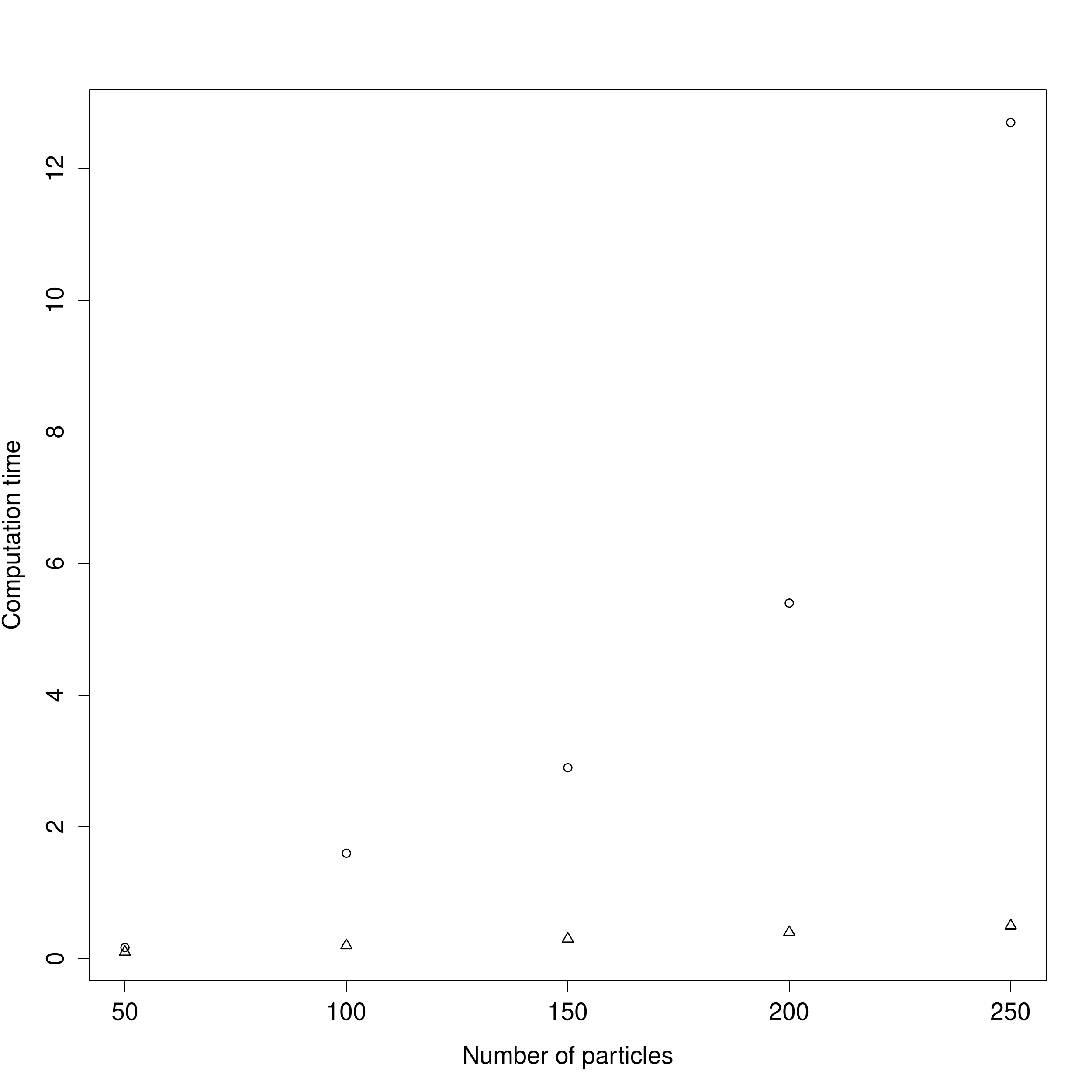}}
\caption{\small Computational time (in minutes) of the standard population ABC of \cite{MPRR12} (triangles) and QMC population
ABC (circles) samplers, for increasing number of particles. The former increases linearly, but the latter does
not.\label{fig:comptime}}
\end{figure}

There are two points to note from this simple experiment:
\begin{enumerate}
\item The QMC version systematically outperforms the basic population ABC sampler, with a variance typically about 30 times smaller for an equal number of particles.
\item The computational time of the QMC version explodes in a non-linear fashion when the number of particles increases, as shown in Figure \ref{fig:comptime}. This is intriguing, and might be due to the cost of generating a QMC sequence, at least in the implementation I used.
\end{enumerate}

There is certainly much more work to be done to adapt these new ideas to an ABC setting. In particular, there is one
situation where QMC point sets could certainly be useful: in ABC, one typically accepts a proposed value $\theta^*$ it
that value leads to simulated data $\bz$ which is close to the observed data $\by$, in the sense that
$d(\by,\bz)<\epsilon$ for some pseudometric $d$ and threshold $\epsilon$. The threshold $\epsilon$ is note always chosen
in advance, but is sometimes a quantile from a fixed number of attempts: there must surely be situation where QMC could
be used to improve the estimator for such flavours of ABC. I was surprised not to find any investigation of this idea in
the literature.

\hyphenation{Post-Script Sprin-ger}

\end{document}